\documentclass[12pt]{iopart}

\usepackage{times}
\usepackage{graphicx}
\input epsf.tex
\def\DESepsf(#1 width #2){\epsfxsize=#2 \epsfbox{#1}}
\def\NPB{{ Nucl. Phys.} B}

\def\PLB{{ Phys. Lett.}  B}
\def\PRL{ Phys. Rev. Lett.}
\def\PRD{{ Phys. Rev.} D}

\def\AP{ Astropart. Phys.}

\begin{document}

\title{SUSY Dark Matter: Closing The Parameter Space}
\author{ R. Arnowitt and B. Dutta\footnote{Department of Physics, University of Regina, Regina
SK, S4S  0A2 Canada}}
\address{Center For Theoretical Physics,
Department of Physics, Texas A\&M University, College
Station, TX 77843-4242}
\begin{abstract}
We consider here the constraints in SUGRA models on the SUSY parameter
space due to current experimental bounds on the light Higgs mass $m_h$, the $b\rightarrow s \gamma$ decay,  
the amount of neutralino cold dark matter $\Omega h^2$,
and the muon magnetic moment. Models with universal soft breaking (mSUGRA)
and non-universal gaugino or Higgs masses at $M_G$ are examined. For mSUGRA,
the $m_h$, $b\rightarrow s \gamma$ and $\Omega h^2$ constraints imply a lower bound on the
gaugino mass of $m_{1/2} \stackrel{>}{\sim}$300GeV implying the gluino and squarks have
mass $\stackrel{>}{\sim}$700GeV, and the neutralino $\stackrel{>}{\sim}$120GeV. The current status of the
Brookhaven muon g - 2 experiment is reviewed, and if the Standard Model
(SM) contribution evaluated using the $e^+ + e^-$ data is correct, a 2$\sigma$
bound on the deviation of experiment from the SM produces an upper bound on
$m_{1/2}$ that eliminates the "focus point" regions of parameter space. Dark
matter (DM) detection cross sections range from $5\times10^{-8}$ pb to
$5\times10^{-10}$
pb which would be accessible to future planned detectors. The SUSY decay
$B_s\rightarrow \mu^+ +\mu^-$ is seen to be accessible to the Tevatron Run 2B with 15
fb$^{-1}$ luminosity for tanbeta $\stackrel{>}{\sim}30$. The most favorable signals of SUSY for
linear colliders are stau pair production and neutralino pair production,
though it will require an 800GeV machine to cover the full parameter space.
Non-universal models can modify some of the above results. Thus a
non-universal (heavier) gluino mass at $M_G$ can significantly reduce the
lower bound constraints of $b\rightarrow s \gamma$ and $m_h$ giving rise to a lighter
SUSY spectrum. A heavier up Higgs mass can open an additional region with
allowed relic density arising from annihilation via the s-channel $Z$ diagram
with an O(10) larger DM detector cross section.

\end{abstract}
% Leave the next line commented out!
% \maketitle

\section{Introduction}

We are now approaching the time when experiments will tell us what the new
physics is that will replace the Standard Model. Neutrino experiments have
already shown the breakdown of the Standard Model with the evidence of
neutrino masses. However, this does not shed light on how the new dynamics
resolves the gauge hierarchy problem, and there are many suggestions for
this: SUSY and SUGRA models, extra dimensions (large and small),
string/M-theory models, etc.

SUSY models all contain a large number of parameters to be determined
finally by experiment. The more phenomena the model applies to, the more
one will be able to limit the parameter space. We consider here SUSY models
with R-parity invariance, since these have the unexpected prediction of the
existence of dark matter \cite{gold,ellis0}. The requirement that the theory predict the
experimentally measured amount  of dark matter already puts strong
constraints on the parameter space. The MSSM with over 100 free parameters
is not very restrictive. At the other extreme is mSUGRA which is consistent
with grand unification (up to now the only direct experimental evidence in
favor of SUSY!) and has only four new parameters and one sign. It's natural
therefore to start with mSUGRA and see how the current experiments restrict
its parameter space, and hence what predictions can be made for future dark
matter experiments and for accelerator experiments at the LHC and NLC. We
will then perturb the mSUGRA framework with non-universal soft breakings
that are consistent with current data, and thus get some idea of how robust
the mSUGRA predictions are.

The current experiments that most strongly restrict the SUSY parameter
space are the following:

(1) The amount of cold dark matter (CDM). Global fits to the CMB and other
data yield $\Omega_{\rm CDM} h^2 = 0.139 \pm0.026$ \cite{turner}, and we take a 2.5$\sigma$ range
around the central value:
\begin{equation}
 0.07 < \Omega_{\rm CDM} h^2 <0.21
\label{om}\end{equation}
The MAP data, due out in the near future, will be able to significantly
narrow this range.

(2) Higgs mass. The LEP lower bound of $m_h > 114.1$ \cite{higgs1} is a significant
constraint for lower $\tan\beta$ (i. e. $\tan\beta \stackrel{<}{\sim}30$), but generally will
become very significant if the bound were to rise by just a few GeV.
Unfortunately, theoretical calculations still have an error of about (2 -
3) GeV, and so we will (conservatively) interpret this experimental bound
to mean that that the theory calculation of $m_h$ should obey $m_h > 111$ GeV.
mSUGRA predicts $m_h \stackrel{<}{\sim}130$ GeV, which could make the light Higgs within
the reach of the Tevatron Run2B.

(3) $b\rightarrow s \gamma$ decay. The CLEO data\cite{bsgamma} has both systematic and
theoretical error, and so we take a relatively broad range around the
central value:
\begin{equation} 1.8\times10^{-4} < B(B \rightarrow X_s \gamma) <
4.5\times10^{-4}
\label{bs}
\end{equation}
The $b\rightarrow s \gamma$ rate is significant for large $\tan\beta$ 
($\tan\beta \stackrel{>}{\sim}30$) where
it produces a lower bound on the gaugino mass $m_{1/2}$. If the lower bound on
the branching ratio were raised, it would increase this lower bound on $m_{1/2}$.

(4) Muon magnetic moment anomaly, $a_\mu$. Since this conference, the
Brookhaven E821 experiment has released new results that have halved the
statistical error with the central value essentially unchanged \cite{BNL}. Also,
new data from CMD-2 and BES has allowed a better evaluation of the SM
contribution and two new theoretical evaluations have appeared \cite{dav,hag}. Using
the $e^+ -e^-$ data to evaluate the hadronic  contribution, both analyses lead
now to a 3$\sigma$ deviation. e.g. \cite{dav}:
\begin{equation}  
\Delta a_\mu = 33.9 (11.2)\times10^{-10}\label{g2}
\end{equation}
If, on the other hand,  the tau decay data (with CVC breaking corrections
included) is used, one finds only a 1.6$\sigma$ discrepancy \cite{dav}, and further,
the two results are statistically inconsistent. In the following, we will
assume that the $e^+ -e^-$ data is correct, and take a 2$\sigma$ band around the
central value of Eq.(3) to see what this implies. However, it may turn out
that the tau data evaluation is correct, under which circumstances, the
current $a_\mu$ data will give no significant constraint.

Future experiments at accelerators will eventually determine if SUSY is
correct. Thus Run2B at the Tevatron should be sensitive to $B_s\rightarrow \mu^+ +\mu^-$
with 15fb$^{-1}$/detector data \cite{bdutta3}, and perhaps distinguish different SUGRA
mediation models \cite{baek}. The LHC can, of course measure SUSY masses and
couplings, e.g. for the gluino up to 2.5 TeV (i.e. $m_{1/2}\stackrel{>}{\sim} 1$ TeV), while the
NLC might observe stau and neutralino production processes.

\section{mSUGRA Model}

The mSUGRA model depends upon four parameters and one sign. It is
convenient to chose these as follows:(i) $m_0$, the scalar soft breaking mass
at the GUT scale $M_G$. (2) $m_{1/2}$, the gaugino mass at $M_G$. [Note that 
$m_{\tilde\chi^0_1}\cong 0$.4 $m_{1/2}$ (where $\tilde\chi^0_1$ is the lightest neutralino), $m_{\tilde\chi^\pm_1} \cong 0.8m_{1/2}$  
(where $\tilde\chi^\pm_1$ is the lightest chargino),
and $m_{\tilde g}\cong2.5m_{1/2}$ (where ${\tilde g}$ is the gluino).] (3) $A_0$ the cubic soft
breaking mass at $M_G$. (4) $\tan\beta = <H_2>/<H_1>$ at the electroweak scale,
where $H_{2,1}$ gives mass to the (up,down) quarks. In addition there is the
sign of $\mu$, the Higgs mixing parameter in the superpotential term $W = \mu
H_1 H_2$. We examine the parameter space over the following range: $m_0 >0$,
$m_{1/2} < 1$TeV ($m_{\tilde g} < 2.5$ TeV), $2 < \tan\beta < 55$, and 
$|A_0| <  4 m_{1/2}$.

For heavy nuclei, the spin independent neutralino-nucleus cross section
dominates for terrestrial dark matter detectors, which allows one to
extract the $\tilde\chi^0_1$ - proton cross section $\sigma_{\tilde\chi^0_1-p}$. (Here
$\tilde\chi^0_1$ is the dark
matter candidate.) The neutralino scattering by
quarks in the nuclear target proceeds through s-channel squark states, and
t-channel Higgs boson states ($h$, $H$). To calculate the relic density of
neutralinos left over after the Big Bang, one needs the neutralino
annihilation amplitudes. The annihilation proceeds through the s-channel Z
and Higgs channels ($h$, $H$, $A$), and t-channel sfermion channels. However, if
a second particle is nearly degenerate with the $\tilde\chi^0_1$, one must include it
in the early universe annihilation processes, which leads to the
co-annihilation phenomena. In SUGRA models this accidental near degeneracy
occurs naturally for the light stau, $\tilde\tau_1$. Co-annihilation then begins at
$m_{1/2}\stackrel{\sim}{=}$ (350-400)GeV and the scalar mass $m_0$ must be raised as $m_{1/2}$
increases to keep the $\tilde\tau_1$ heavier than the $\tilde\chi^0_1$.

The neutralino -proton cross section has the following general behavior:
$\sigma_{\tilde\chi^0_1-p}$ increases with increasing $\tan\beta$, and decreases with increasing
$m_{1/2}$, $m_0$. The maximum  value of $\sigma_{\tilde\chi^0_1-p}$ then generally occurs at large
$\tan\beta$ and small $m_{1/2}$, $m_0$.

One starts the analysis at $M_G$ and uses the renormalization group equations
(RGE) to obtain predictions at the electroweak scale. In carrying out these
calculation, it is necessary to include a number of corrections and we
list some of these here:
(1) We use two loop gauge and one loop Yukawa RGE in running from $M_G$ to the
electroweak scale $M_{\rm EW}$, and three loop QCD RGE below $M_{\rm EW}$  for light quark
contributions.
(2) Two loop and pole mass corrections are included in the calculation of $m_h$.
(3) One loop correction to $m_b$ and $m_\tau$ are included\cite{rattazi}.
(5) All stau-neutralino co-annihilation channels are included in the relic
density calculation \cite{bdutta,ellis,gomez}. (Chargino-neutralino co-annihilation does
not occur for $m_{1/2} < 1$ TeV.) Large $\tan\beta$ NLO SUSY corrections to $b\rightarrow
s\gamma$ are included\cite{degrassi,carena2}.
We do not include Yukawa unifications or proton decay constraints as these
depend sensitively on post-GUT physics, about which little is known.

Figs. 1 and 2 illustrate the allowed regions in the $m_0 - m_{1/2}$ plane for
$\tan\beta = 10$ and $\tan\beta = 50$ for $A_0 = 0$, $\mu >0$. We note for $\tan\beta = 10$,
that the lower bound on $m_{1/2}$ is set by the lower bound on the Higgs mass.
The narrowness of the dark matter allowed band does not imply any fine
tuning. Thus the lower edge is determined from the co-annihilation effect
so that the $\tilde\tau_1$, which is nearly degenerate with the $\tilde\chi^0_1$ does not cause
too much early universe annihilation (violating the lower bound of Eq(1)),
while above the upper bound of the band, too little early universe
annihilation occurs (violating the upper bound of Eq. (1)). Thus the MAP
(and Planck) satellite data will narrow this band further, with more
accurate determinations of $\Omega_{\rm CDM}$.  At $\tan\beta = 50$, the allowed dark
matter band expands at low $m_{1/2}$ due to the fact that the A Higgs becomes
light, allowing additional early universe annihilation, and this is
compensated by raising $m_0$. Note here it is the $b\rightarrow s \gamma$ branching ratio
that give the lower cut off in $m_{1/2}$. The vertical lines through the dark
matter allowed bands represent the values of $\sigma_{\tilde\chi^0_1-p}$. We see that the
cross sections range from $5\times10^{-8}$ pb to $10^{-9}$ pb. The $\mu < 0$
possibility (not shown) allows for regions of much smaller dark matter cross sections
\cite{ellis3,bdutta}. However, if the SUSY contribution to $a_\mu$ is indeed positive
(which is the case for the $e^+ -e^-$ evaluation of the SM contribution), then
the $\mu < 0$ possibility is eliminated \cite{nano,chat}.

\begin{figure}
\centerline{ \DESepsf(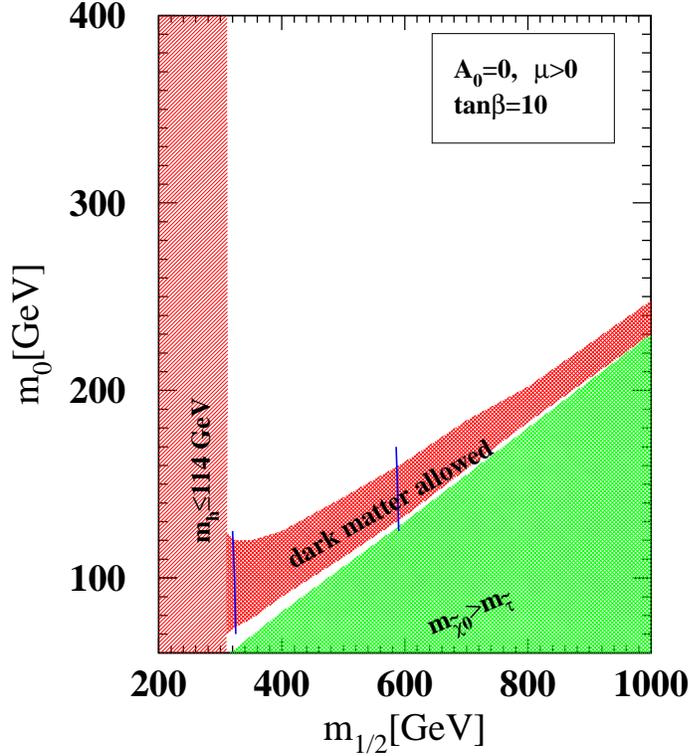 width 8 cm) }
\caption {\label{fig1}  Allowed regions in $m0 - m_{1/2}$ parameter space for $\tan\beta = 10$, $A_0 =
0$, $\mu >0$. The left vertical line corresponds to 
$\sigma_{\tilde\chi^0_1-p} = 5\times10^{-9}$ pb
and the right to $10^{-9}$ pb.} 
\end{figure}
					    
\begin{figure}
\centerline{ \DESepsf(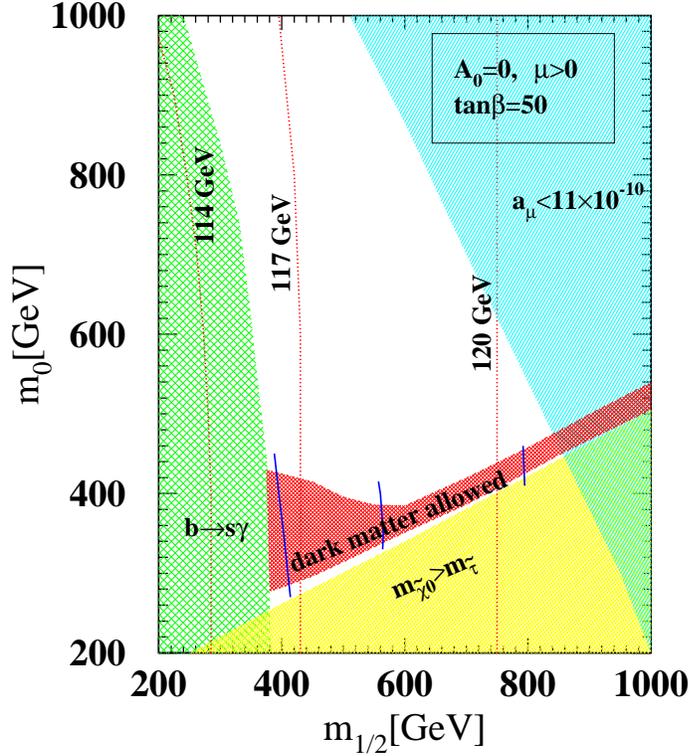 width 8 cm) }
\caption {\label{fig2}  Same as Fig. 1 with $\tan\beta = 50$, and 
$\sigma_{\tilde\chi^0_1-p} =5\times 10^{-8}$
(left) and $2\times10^{-9}$ (right).} 
\end{figure}

\section{The Muon Magnetic Moment Anomaly}

As discussed in the introduction, it is still unclear if the current
measurements of the muon magnetic moment indicate a deviation with the
Standard Model prediction. In this section we will assume that the analysis
based on the $e^+ -e^-$ data, which yields the 3$\sigma$ deviation of Eq.(3), is
correct, in order to see what restriction on the SUSY parameter space this
possibility allows. In supersymmetry, one expects a contribution to the
muon magnetic moment of characteristic size $10\times10^{-10}$ arising from
chargino-sneutrino and neutralino-smuon loops (with the photon attached to
any charged line). If the final result requires $a_\mu^{\rm SUSY}$ to be much
smaller than $10\times10^{-10}$, it would mean that the SUSY squark and gluino
masses are in the TeV domain, while a value of 
$a_\mu^{\rm SUSY}\stackrel{>}{\sim} 40\times10^{-10}$
would exclude mSUGRA \cite{bdutta2}. Thus the allowable range for $a_\mu^{\rm SUSY}$ is
already somewhat restricted. Fig.3 exhibits the effect of the muon magnetic
moment anomaly for the case of $\tan\beta = 40$, $A_0 = 0$, $\mu > 0$. The shaded
region in the upper right would be the part of the parameter space excluded
at the 2$\sigma$ level (assuming the $e^+ -e^-$ evaluation of the SM). Combined
with the dark matter constraint, one sees that the $a_\mu$ constraint give
rise to an upper bound  $m_{1/2} \stackrel{<}{\sim} 800$GeV or $m_{\tilde \chi^0_1} <
320$ GeV, $m_{\tilde \chi^{\pm}_1} < 640$ GeV
and $m_{\tilde g} < 2$ TeV. The $b\rightarrow s\gamma$ constraint gives a lower bound of 
$m_{1/2}>350$GeV, and thus the parameter space is constrained at both ends. Note also
that it would not take a significantly larger value then the central value
of $33\times10^{-10}$ to exclude the full parameter space for this value of
$\tan\beta$. Thus a resolution of the Standard Model contribution to $a_\mu$ is of
great importance for SUSY models.

 \begin{figure}
\centerline{ \DESepsf(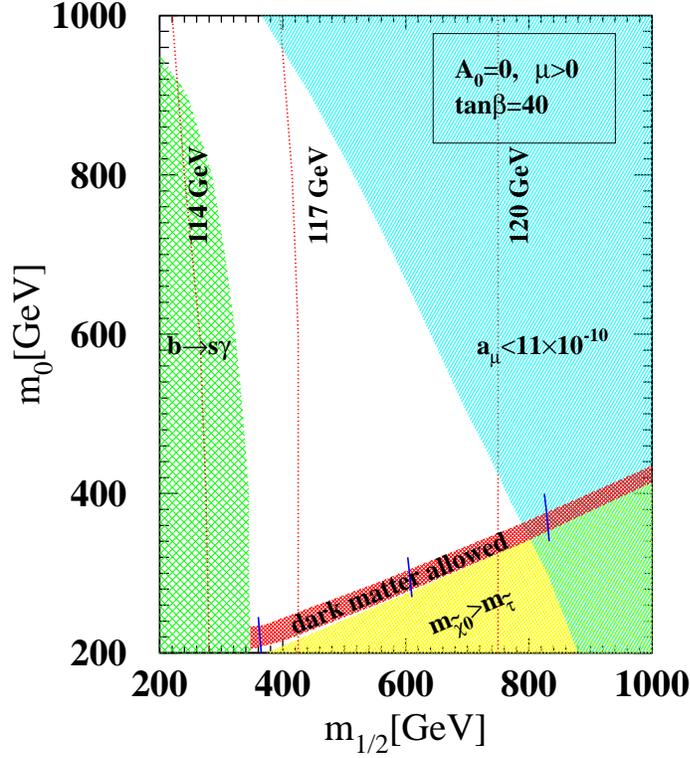 width 8 cm) }
\caption {\label{fig3}  Allowed region in the $m_0 - m_{1/2}$ plane for $\tan\beta = 40$, $A_0 = 0$,
$\mu > 0$. The shaded upper region is excluded at the 2$\sigma$ level for $a_\mu$
obeying Eq.(3).} 
\end{figure}

\section{The $B_s\rightarrow\mu^+ + \mu^-$ Decay}

The $B_s\rightarrow\mu^+ + \mu^-$ decay \cite{bobeth,bdutta3,dedes,tata} offers another window for
investigating the mSUGRA parameter space. The Standard Model predicts a
branching ratio that is quite small i. e. B[$B_s\rightarrow\mu^+ + \mu^-$] =
$(3.1\pm1.4)\times10^{-9}$. However, the SUSY contribution can become quite large
for large $\tan\beta$. An example of one of the important graphs for this
process is shown in Fig. 4, where the amplitude grows as $(\tan\beta)^3$, and
hence the rate grows as $(\tan\beta)^6$.  This process has become interesting
because it appears possible to observe it at Run2B of the Tevatron \cite{bdutta3}. A
set of cuts eliminating the background (e.g. gluon splitting of $g\rightarrow b {\bar b}$
and non-b background) exits and leads to a sensitivity for CDF for a
luminosity of 15fb$^{-1}$ of
\begin{equation}
B[B_s\rightarrow\mu^+ + \mu^-] \stackrel{>}{\sim}1.2\times 10^{-8}
\label{Bsmumu}
\end{equation}
(and if a similar sensitivity can be obtained by D0 the combined
sensitivity would be $0.65\times 10^{-8}$ ).  Fig. 5 shows  the branching ratio for
different $\tan\beta$  for the case $m_0 = 300$GeV, $A_0 = 0$, $\mu > 0$, and Fig. 6
shows the expected limit on the branching ratio observable by CDF as a
function of luminosity. We see that with 15 fb$^{-1}$, CDF should be
sensitive to this decay for $\tan\beta \stackrel{>}{\sim}30$, and in fact too large a
branching ratio could exclude mSUGRA with 2 fb$^{-1}$.

 \begin{figure}\vspace{-0cm}
\centerline{ \DESepsf(figadkm.epsf width 8 cm) }
\caption {\label{fig4}  A leading diagram for the decay $B_s\rightarrow\mu^+ + \mu^-$. The vertices with
the heavy dot are proportional to $\tan\beta$.} 
\end{figure}

\begin{figure}
 \centerline{ \DESepsf(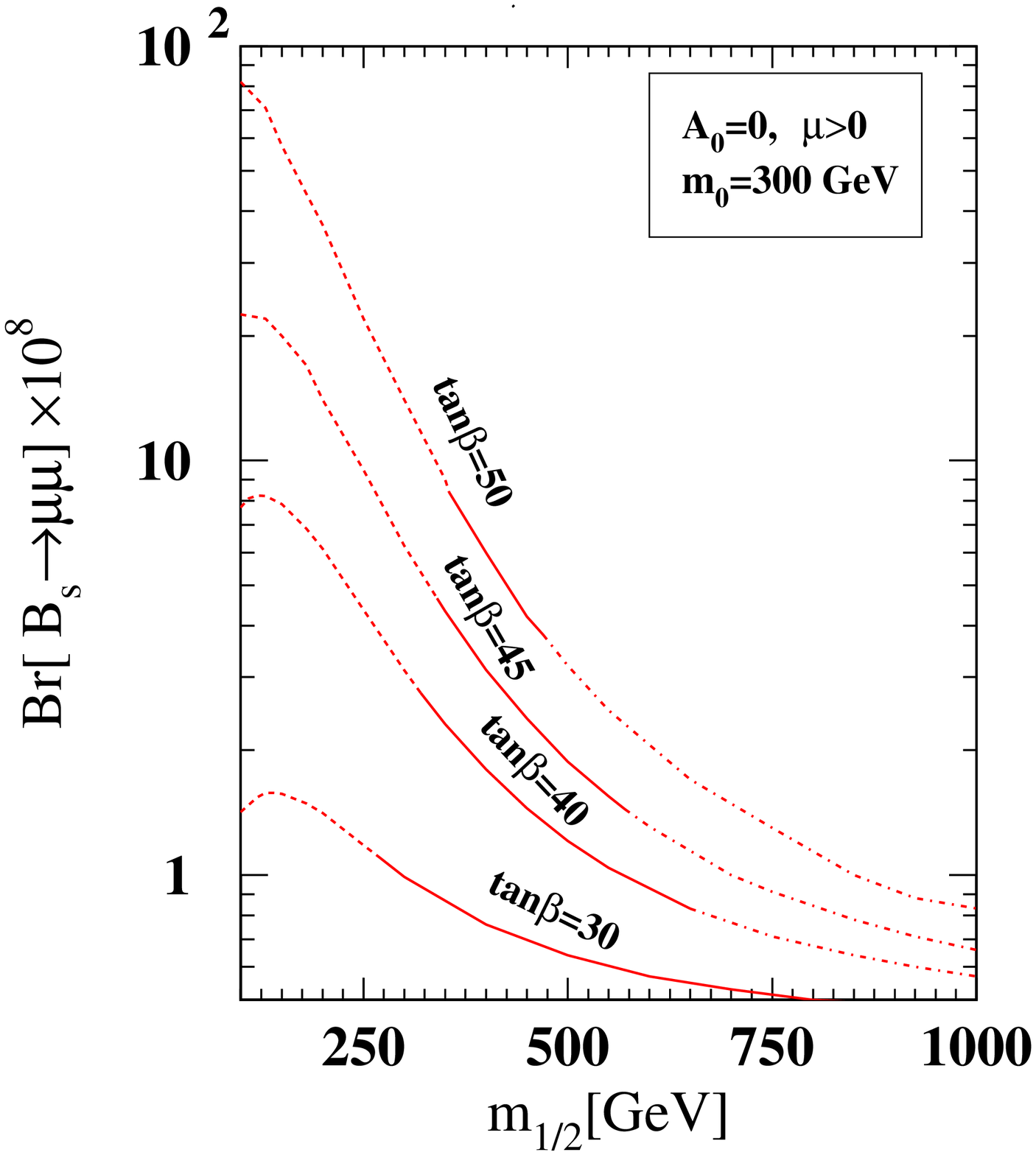 width 8 cm) } 
\caption {\label{fig5}  The branching ratio of $B_s\rightarrow\mu^+ + \mu^-$ as a function of
$m_{1/2}$ for
different values of $\tan\beta$[9].}
\end{figure}

\section{NLC Reach}

We consider here two possible accelerators at energies of 500GeV and 800GeV to
examine what part of the SUSY spectrum would be available to linear
colliders. We've seen that the $m_h$ and $b \rightarrow s \gamma$ bounds already means
that $m_{1/2}\stackrel{>}{\sim}(350 - 400)$ GeV, and so for mSUGRA, gluinos and squarks would
generally be beyond the reach of such machines. The most favorable SUSY
signals are then
\begin{eqnarray}
            e^+ + e^- &\rightarrow& \tilde\chi^0_2 + \tilde\chi^0_1 \rightarrow  
	    (l^+ + l^- +\tilde\chi^0_1) +\tilde\chi^0_1 \\\nonumber 
	    e^+ + e^- &\rightarrow& \tilde\tau_1^+ + \tilde\tau_1^- \rightarrow 
( \tau +\tilde\chi^0_1) + ( \tau +\tilde\chi^0_1)
\end{eqnarray}
where $l^{\pm}$ is any charged lepton. Since in mSUGRA $m_{\tilde\chi^0_2} \simeq 2m_{\tilde\chi^0_1}$ one has
for the kinematic mass reach $1/2m_{\tilde\chi^0_2} \simeq m_{\tilde\chi^0_1} \stackrel{<}{\sim}165 (265)$ GeV for 
$\sqrt s
= 500 (800)$ GeV, and $m_{\tilde\tau_1}\stackrel{<}{\sim}250 (400)$ GeV for $\sqrt s = 500 (800)$ GeV.
In general the stau is the lightest slpeton and so will be the major
signal, as the selectron and smuon will not be accessible for most of the
parameter space. There are a number of backgrounds however, and appropriate
cuts have to be arranged to suppress them. This is currently being studied
[bdutta4].

\begin{figure}\vspace{-0cm}
 \centerline{ \DESepsf(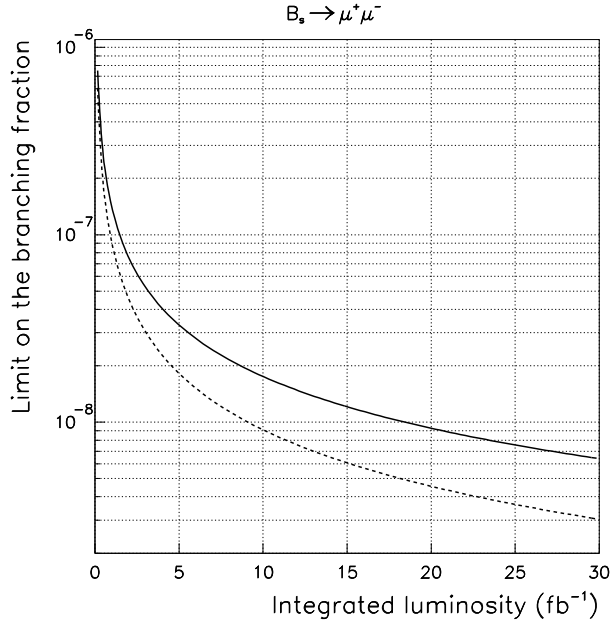 width 8 cm) } 
\caption {\label{fig6}  The branching ratio of $B_s\rightarrow\mu^+ + \mu^-$ as a function of
$m_{1/2}$ for
different values of $\tan\beta$.[9]}
\end{figure}

\section{ The mSUGRA Parameter Space}

We now summarize the effects on the mSUGRA parameter space of the
constraints from $m_h$, $b\rightarrow s\gamma$, dark matter, and $a_\mu$ (assuming Eq. (3)
is valid). We will also see what effect might be expected from measurement
of the  $B_s\rightarrow \mu \mu$ decay at the Tevatron, and examine the NLC reach for
SUSY. Fig. 7 exhibits these effects for $\tan\beta=$10, $A_0 =0$, $\mu >0$. 
We see
that for low $\tan\beta$, the $a_\mu$ constraint eliminates most of the parameter
space, and the NLC with $\sqrt{s} = 500$ GeV would be able to scan the full
remaining space with either the stau or neutralino signal. The
$\sigma_{\tilde\chi^0_1-p}$
cross sections are small, but of the size that future experiments hope to
achieve (e.g. GENIUS, Cryoarray, ZEPLIN  IV, CUORE).Of the two NLC signals,
the $\tilde\chi^0_2 - \tilde\chi^0_1$ is sensitive to large $m_0$, while the 
$\tilde\tau_1 -\tilde\tau_1$ is sensitive
to relatively large $m_{1/2}$. Due to the dark matter constraint, the latter
appears to be the more important signal.

\begin{figure}\vspace{-0cm}
 \centerline{ \DESepsf(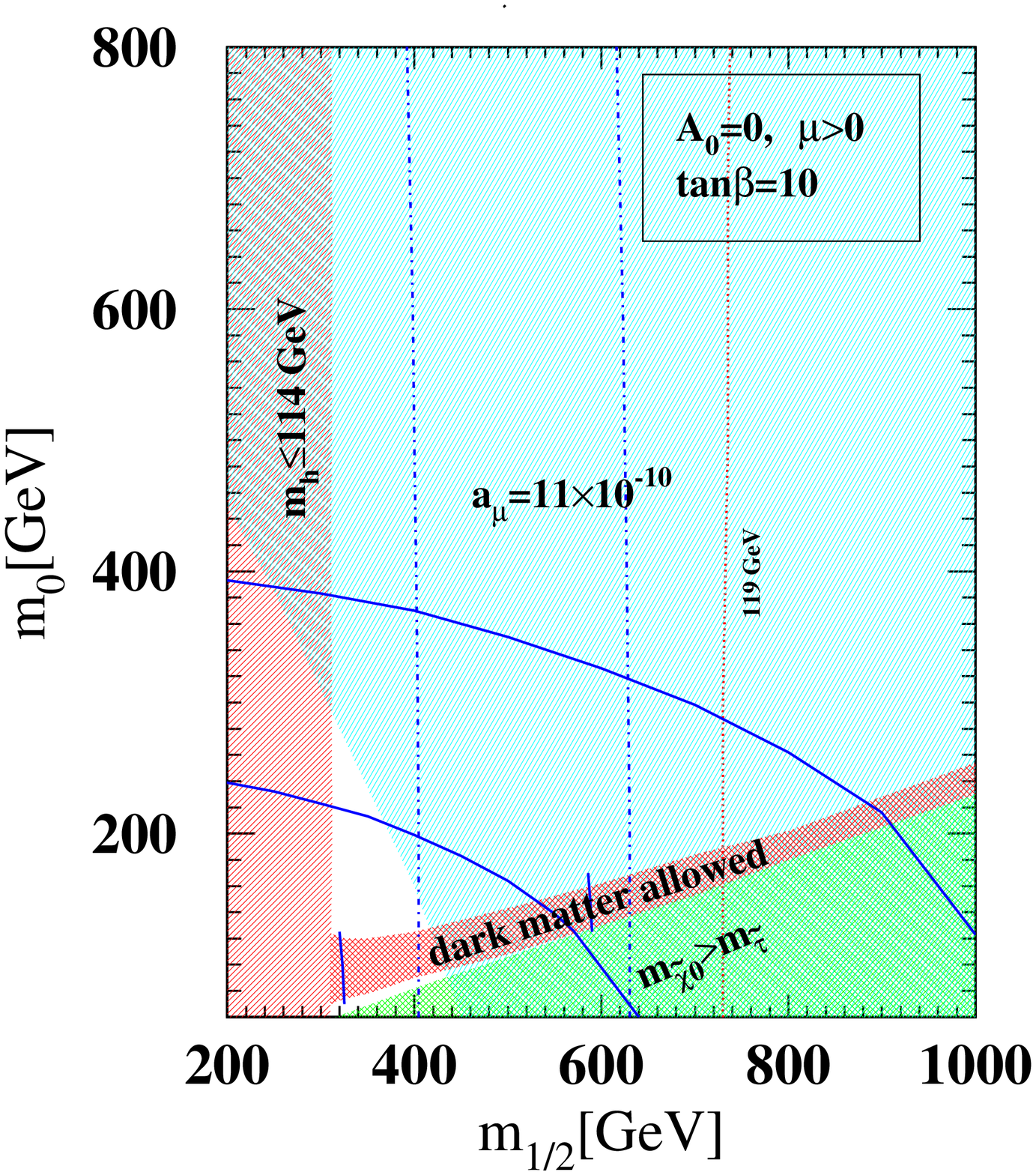 width 8 cm) } 
\caption {\label{fig7}  Allowed region in the $m_0- m_{1/2}$ plane for $\tan\beta = 10$, $A_0 =
0$, $\mu >
0$. The shaded upper region is the $2\sigma$ $a_\mu$ bound. The vertical dash-dot
lines are the kinematical bound at a linear collider for the $\tilde\chi^0_1 -
\tilde\chi^0_2$ signal, the left line for a 500 GeV machine and the right for 800 GeV.
(Parameter space to left of these lines would be observable if the
kinematical reach could be achieved.) The curved lines are similarly for
the $\tilde\tau_1-\tilde\tau_1$ signal at a linear collider. The solid vertical lines through
the dark matter allowed band give the values of $\sigma_{\tilde\chi^0_1-p}$ i.e. 
$5\times10^{-9}$pb
for the left line and $1\times10^{-9}$ for the right line.}
\end{figure}

Fig. 8 shows similar information for $\tan\beta = 40$. For larger $\tan\beta$, the
possibility of observing the $B_s\rightarrow \mu \mu$ signal at the Tevatron becomes
significant, and assuming the combined CDF and D0 data were available, it
could cover the full parameter space for an $a_\mu$ anomaly $> 10\times 10^{-10}$. The
NLC (500GeV)  tau-tau signal would now cover only about one half the
allowed parameter space, while for 800GeV the full parameter space could be
examined.
\begin{figure}\vspace{-2cm}
 \centerline{ \DESepsf(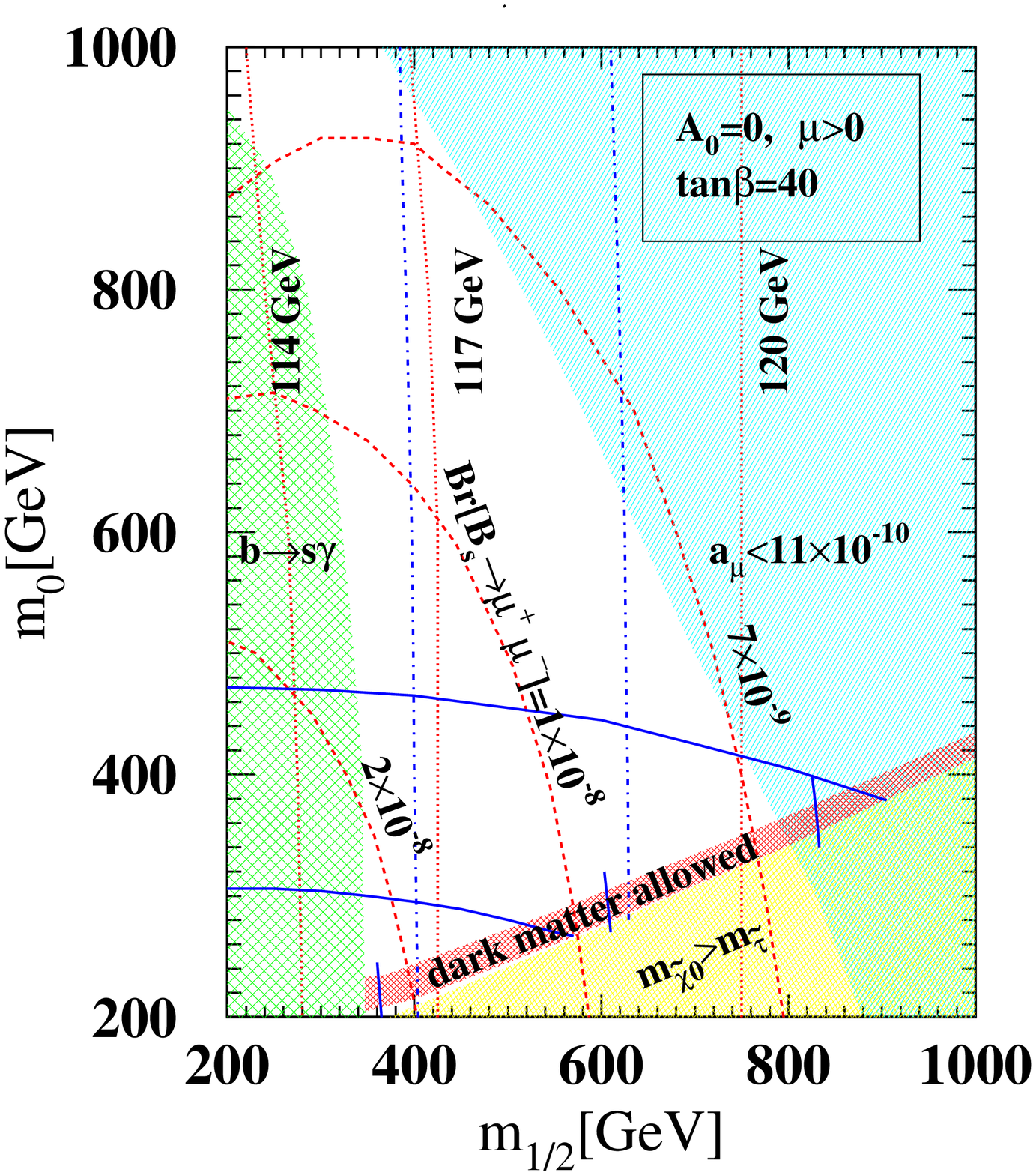 width 8 cm) } 
\caption {\label{fig8} Same as Fig. 7 with dashed lines giving $B_s\rightarrow \mu \mu$ Tevatron
reaches. The left solid dark line corresponds to
$\sigma_{\tilde\chi^0_1-p}=3\times10^{-8}$ and
the extreme right to $1\times10^{-9}$.}
\end{figure}

For very large tanbeta, a new feature of a ``bulge" occurs in the dark
matter channel at low $m_{1/2}$ in the allowed dark matter channel. This is due
to the fact the heavy Higgs A and H become light, and allow a more rapid
annihilation in the early universe through the A and H s-channel poles.
Figs. 9 and 10 for $\tan\beta =$ 50 and 55 exhibit this feature. (We note in
general that the results for $\tan\beta\geq 45$ become very sensitive to the
precise values of $m_t$ and $m_b$. We are here using the central values of 
$m_t(pole) =175$ GeV and $m_b(m_b) = 4.25$GeV, but the figures would change significantly if one
were to go one or two $\sigma$ away from the mean.) We see from the figures
that the $\tilde\tau_1-\tilde\tau_1$ LC signal would be able to cover the full parameter space
of $\tan\beta = 50$ for an 800 GeV collider, but not quite cover it all for
$\tan\beta = 55$. Of course, for these large tanbeta, the $B_s\rightarrow \mu \mu$ signal
is greatly enhanced, as is the dark matter detector cross sections.

\begin{figure}\vspace{-0cm}
 \centerline{ \DESepsf(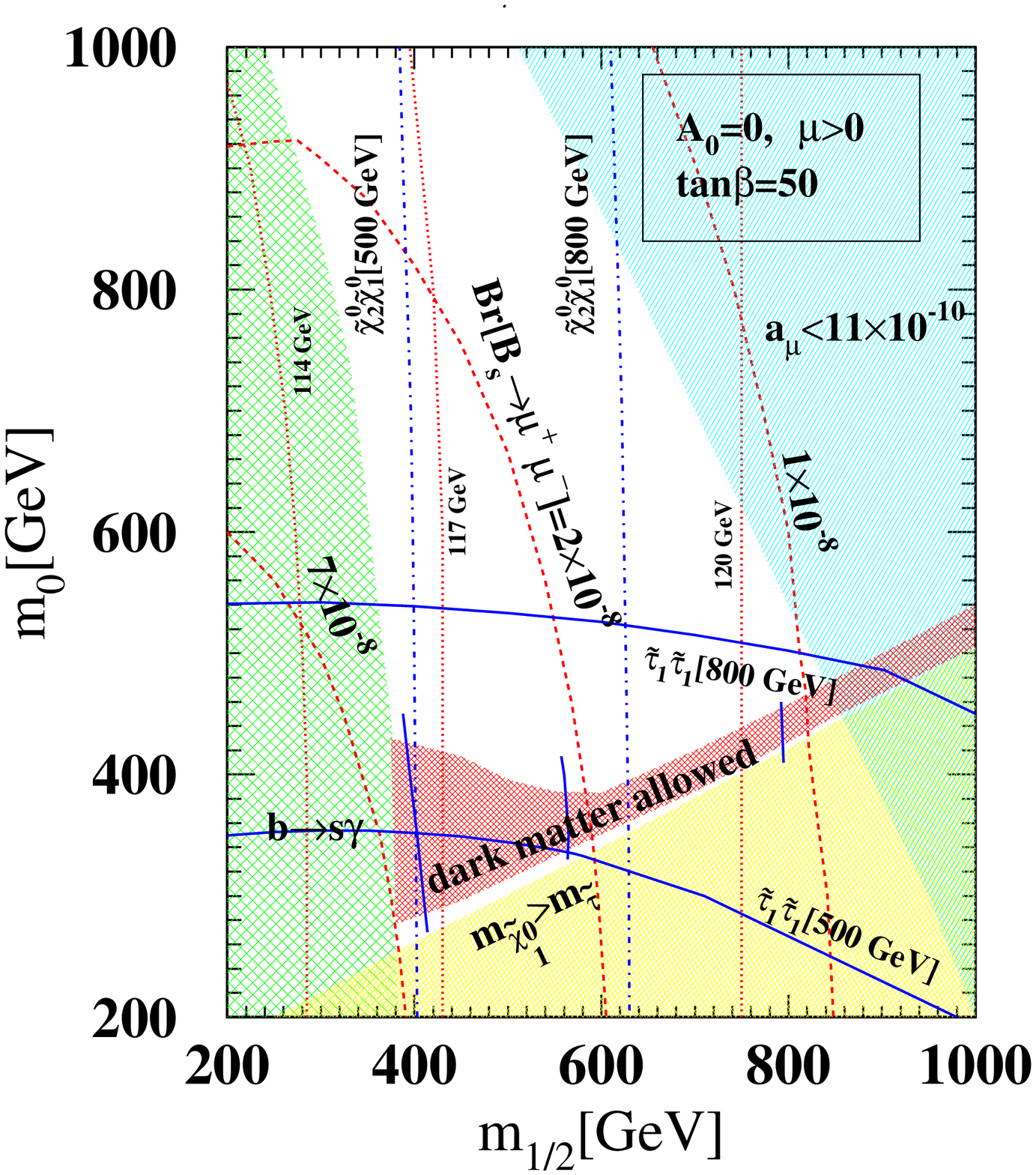 width 8 cm) } 
\caption {\label{fig9}Same as Fig. 8 for $\tan\beta = 50$, with the left solid line
corresponding to $\sigma_{\tilde\chi^0_1-p}= 5\times 10^{-8}$pb and the right to 
$2\times10^{-8}$pb.}
\end{figure}

\begin{figure}\vspace{-0cm}
 \centerline{ \DESepsf(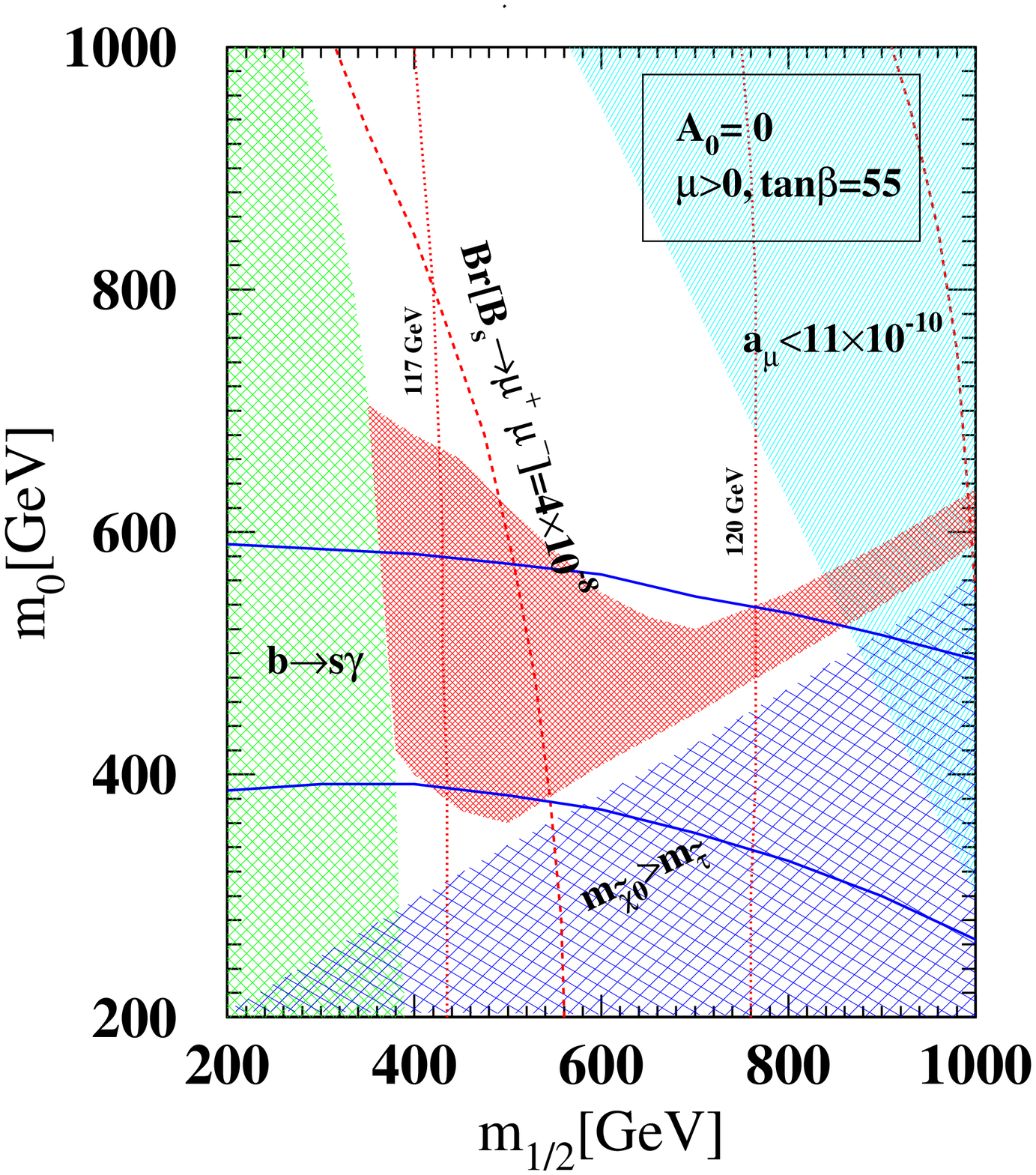 width 8 cm) } 
\caption {\label{fig10}Same as Fig. 8 for $\tan\beta = 55$.}
\end{figure}

\section{ Non-universal Models}

Our previous discussion has been within the framework of mSUGRA with
universal soft breaking occurring. It is interesting to investigate what
aspects of these results would survive in the presence of some
non-universal soft breaking. One peculiar feature of mSUGRA is the narrow
dark matter allowed bands due in part to the ``accidental" near degeneracy
of the $\tilde\tau_1$ and the $\tilde\chi^0_1$. This phenomena would be maintained even if the
gaugino masses were non-universal at $M_G$, since $m_{\tilde\tau_1}^2 - m_{\tilde\chi^0_1}^2$ depends
mainly on $\tilde m_1$. Thus this type of co-annihilation effect is generic. A
second feature of mSUGRA, that the combined effects of the $m_h$ and $b\rightarrow s\gamma$
experimental bounds put a lower limit $m_{1/2} > (300-400)$ GeV for all
tanbeta, is however sensitive to the mSUGRA assumption of universal gaugino
masses at $M_G$. Thus if one assumes gluino breaking of this universality at $M_G$,

\begin{equation}  
\tilde m_1 = \tilde m_2 = \tilde m_3(1 +
\tilde\delta_3),\label{gl}
\end{equation}  
then a positive $\tilde\delta_3$ raises the stop mass and effects strikingly both
the $b\rightarrow s\gamma$ branching ratio and the value of $m_h$. This is illustrated in
Fig. 11 for the case of $\tilde\delta_3$= 1, $\tan\beta =50$, $A_0=0$, $\mu >0$. One sees
that both the $b\rightarrow s\gamma$ and $m_h$ constraint is moved strongly towards
lower $m_{1/2}$, and now the lower bound on $m_{1/2}$ is only $m_{1/2}\stackrel{>}{\sim} 190$ GeV. The
Higgs lines are also moved and lie approximately 150 GeV  lower in $m_{1/2}$.
The $a_{\mu}$ bound, however, becomes somewhat more constraining.

A second non-universality that might naturally arise is with the scalar
soft breaking masses. Thus while flavor changing neutral current constraint
require squark masses to be nearly degenerate at the GUT scale, there is little theoretical
reason to require that the Higgs masses be degenerate. We consider here the
simple model where at $M_G$
\begin{equation}  
m_{H_1}^2 = m_0^2 ( 1 + \delta_1);\,\,   m_{H_2}^2 =
m_0^2  (1 + \delta_2)      \label{hl}
\end{equation}
where $m_0$ is the universal squark and slepton masses.  While this
introduces two new parameters into the model, one may qualitatively
understand the effects of $\delta_{1,2}$. In SUGRA models, the $\mu$ parameter
governs much of the physics, and if $\mu^2$ decreases (increases), then the
Higgsino content of the neutralino $\tilde\chi^0_1$ increases (decreases). The RGE
with radiative breaking shows that $\mu^2$ is indeed sensitive to $\delta_2$ (but
only slightly sensitive to $\delta_1$). For $\delta_2$ positive, $\mu^2$ decreases,
and this produces two effects: (1) it increases the dark matter detection
cross section $\sigma_{\tilde\chi^0_1-p}$ since this quantity depends on the interference
between the Higgsino and gaugino parts of the neutralino. (2) It increases
the $\tilde\chi^0_1 - \tilde\chi^0_1 -Z$ coupling, allowing more early universe annihilation of
the neutralinos through the Z s-channel. This then opens another channel of
allowed relic dark matter in the $m_0- m_{1/2}$ plane at relatively low $m_{1/2}$ but
high $m_0$ (so that too much annihilation doesn't occur). These effects are
seen in Figs. 12 and 13. Fig. 12 shows the new $Z$-channel allowed dark
matter region at high $m_0$, along with the usual stau co-annihilation narrow
band at low $m_0$ for $\tan\beta = 40$. Fig. 13 shows the corresponding values of
$\sigma_{\tilde\chi^0_1-p}$, the upper dashed band corresponding to the $Z$-channel of 
Fig. 13.
One sees that the detector cross section for this possibility can be a
factor of ten or more larger than the usual stau band.

\begin{figure}\vspace{-0cm}
 \centerline{ \DESepsf( 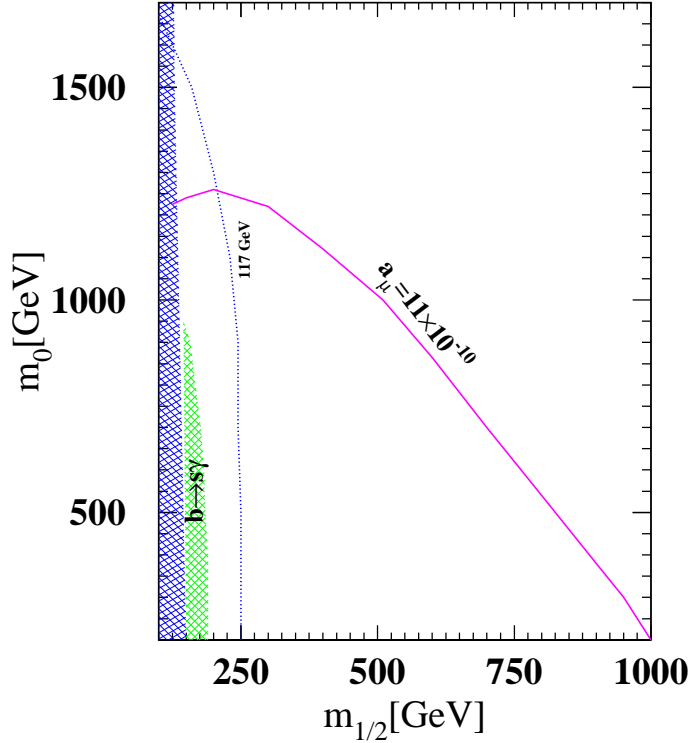 width 8 cm) } 
\caption {\label{fig11} The $b\rightarrow s\gamma$, $m_h$ and $a_{\mu}$ constraints for $\tilde\delta_3 = 1$, 
$\tan\beta =50$, $A_0 =0$, $\mu > 0$.}
\end{figure}

\begin{figure}\vspace{-0cm}
 \centerline{ \DESepsf( 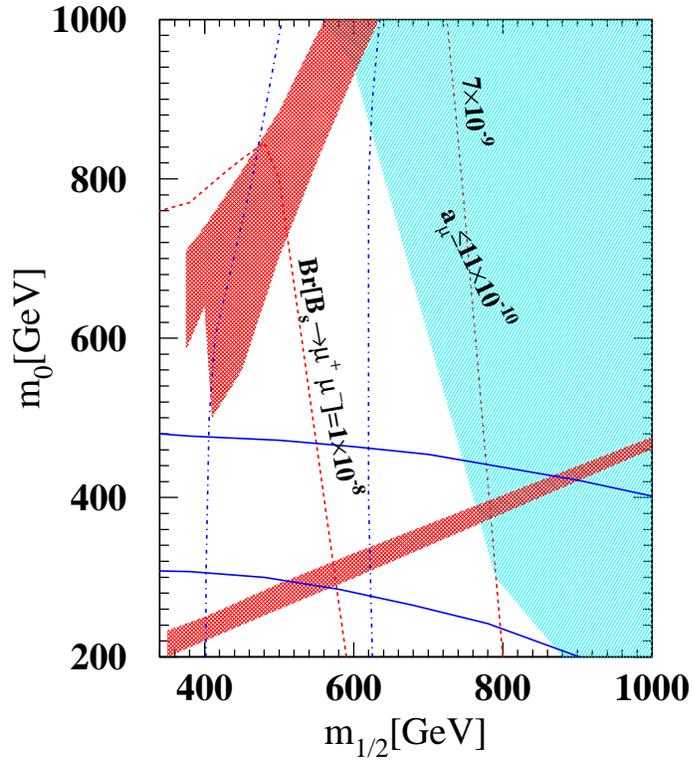 width 8 cm) } 
\caption {\label{fig12} Allowed region in the $m_0- m_{1/2}$ plane for $\delta_2 = 1$, 
$\tan\beta =40$, $A_0 = m_{1/2}$, $\mu > 0$. The lower narrow band is the ususal stau
co-annihilation band, and the upper band is due to the $Z$-channel annihilation.}
\end{figure}

\begin{figure}
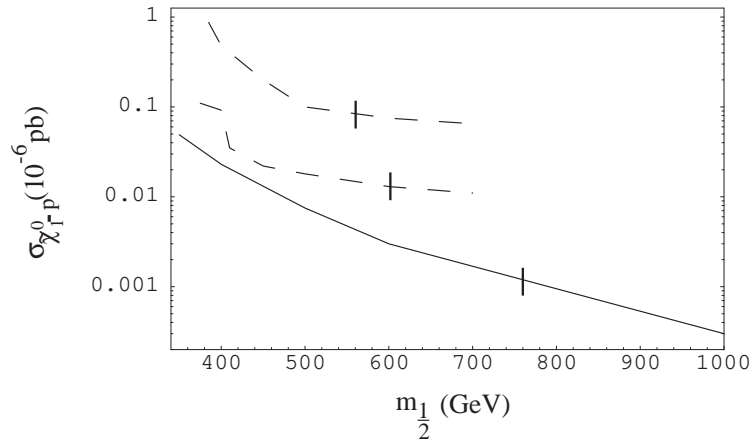
\vspace{-0cm}
 \centerline{ \DESepsf(adhs4.epsf width 10 cm) } 
\caption {\label{fig13}Dark matter detection cross section for the two allowed regions in
the $m_0- m_{1/2}$ of Fig. 12. The upper dashed band is for the Z-cannel
annihilation, and the lower line is for the $\tilde\tau_1 - \tilde\chi^0_1$ co-annihilation
channel. [12]}
\end{figure}

\section{ Conclusions and Summary}

We have examined here for SUGRA models of neutralino dark matter, the
constraints on the SUSY parameter space that arise from current
experiments, and have considered models both with universal soft breaking
(mSUGRA) and with non-universal soft breaking.

For mSUGRA, co-annihilation effects and the current bounds on the relic
density generally restrict the allowed parameters to be in a narrow band in
the $m_0- m_{1/2}$ plane with relatively low $m_0$ (except in the low $m_{1/2}$ region
where for very large $\tan\beta$ a ``bulge" can exist). If the SUSY contribution
to the muon magnetic moment anomaly is $\stackrel{>}{\sim} 10\times 10^{-10}$ (which is the current
2$\sigma$ bound for the $e^+ -e^-$ data analysis of the SM contribution\cite{dav,hag}),
then the "focus point" region \cite{feng} of very large $m_0$ is also eliminated. The
current bounds on $m_h$ and $b \rightarrow s\gamma$ further produce a lower bound on $m_{1/2}$
of $m_{1/2}\stackrel{>}{\sim}$ (300- 350) GeV over the full tanbeta domain. This implies that
$m_{\tilde\chi^0_1} \stackrel{>}{\sim} $120 GeV,  $M_{\tilde\chi^{\pm}_1} \stackrel{>}{\sim}$ 240 GeV  and $m_{\tilde g}\stackrel{>}{\sim}$ 750 GeV. These mass
bounds are larger than what exists now from accelerators. In general we
find that dark matter direct detection neutralino -proton cross sections
range from $\sim5\times10^{-8}$ to $\sim5\times10^{10}$, which are mostly within the range of
sensitivity of future planned dark matter detectors. The $B_s\rightarrow \mu \mu$ decay
will be accessible to the Tevatron over much of the parameter space for
tanbeta $\stackrel{>}{\sim}30$ with 15 fb$^{-1}$ of luminosity. The accessibility of
the  $\tilde\tau_1-\tilde\tau_1$ and $\tilde\chi^0_1 -\tilde\chi^0_2$ SUSY signals for the NLC was examined.

To see how robust the above results are, we examined two types of
non-universal soft breaking models: those with non-universal gaugino masses
at the GUT scale, and those with non-universal Higgs masses. In general for
these types of models, the co-annihilation effects still remain. However,
an increased non-universal gluino mass can significantly reduce the lower
bound constraints on $m_{1/2}$ arising from the $b\rightarrow s\gamma$ and $m_h$ constraints,
making these constraints less significant. ( The $a_\mu$ constraint is
simultaneously somewhat strengthened). Non-universal Higgs soft breaking
masses can also produce some striking effects. Thus an increase of the $H_2$
mass at the GUT scale increases the Higgsino part of the neutralino, which
then increases the $\tilde\chi^0_1 - \tilde\chi^0_1 - Z$ coupling and opens a new  $Z$ s-channel
annihilation region at low $m_{1/2}$ and high $m_0$ with acceptable amount of relic
dark matter. In this region the $\tilde\chi^0_1-p$ detection cross section is a factor
of 10 or more larger than that from the $\tilde\tau_1$ co--annihilation region,
making it accessible to the next round of dark matter detector experiments.
However, SUSY signals in such regions would be difficult to detect for a
500 GeV NLC.

\end{document}